\documentclass[twocolumn,article,amsmath,amssymb]{revtex4}
\usepackage{graphicx}
\usepackage{graphics}
\usepackage{dcolumn}
\usepackage{bm}
\usepackage{units}
\usepackage{braket}
\usepackage{soul}
\usepackage{color}
\usepackage{amsmath}
\usepackage{xfrac}
\usepackage{caption}
\usepackage{subcaption}
\usepackage{epstopdf}
\usepackage{physics}
\usepackage{color}

\makeatletter
\newenvironment{figurehere}
{\def\@captype{figure}}
{}
\makeatother

\begin{document}

\def\e{\mathop{\rm \mbox{{\Large e}}}\nolimits}
\def\im{\mathop{\rm \od{\iota}}\nolimits}
\newcommand{\ts}[1]{\textstyle #1}
\newcommand{\bn}[1]{\mbox{\boldmath $#1$}}
\newcommand{\bc}{\begin{center}}
\newcommand{\ec}{\end{center}}
\newcommand{\be}{\begin{equation}}
\newcommand{\ee}{\end{equation}}
\newcommand{\bea}{\begin{eqnarray}}
\newcommand{\eea}{\end{eqnarray}}
\newcommand{\ba}{\begin{array}}
\newcommand{\ea}{\end{array}}
\newcommand{\red}[1]{\textcolor{black}{#1}}

\title{Role of non-extensivity $q$-parameter in tectonic seismic forecasting}

\author{O. Sotolongo-Costa\footnote{(O. Sotolongo-Costa): osotolongo@gmail.mx} and M. E. Mora-Ramos\footnote{(M. E. Mora-Ramos): memora@uaem.mx}}

\affiliation{Centro de Investigación en Ciencias-IICBA, Universidad Autónoma del Estado de Morelos, Av. Universidad 1001, CP 62209 Cuernavaca, Morelos, México}

\begin{abstract} 
By writing total Tsallis entropy as a function of non-extensivity $q$-parameter withing the fragment-asperity model for earthquakes, a critical range of values is identified: $1.4\lesssim q\lesssim 1.8$. It comes directly from constructing the non-extensive entropy with the assumption that the energy of stress-bearing interactions, which the probability distribution depends on, is proportional to the surface of contact. Such interval of $q$-values corresponds to the strong variation of entropy and contains the most of reported results for this parameter determined for main-shocks around the world in recent decades, indicating the role of $q$ as a seismic risk factor. Although this knowledge is clearly not enough to elaborate a procedure to predict the occurrence of intense and devastating tectonic earthquakes, it may serve as a relevant element of consideration.

\end{abstract}

\maketitle

\section{Introduction}\label{intro}

Earthquakes result from the sudden release of accumulated strain along tectonic faults, a process governed by stick-slip dynamics and heterogeneous stress distributions. The inherent complexity of fault systems—characterized by fractal geometries, scale-invariant stress correlations, and nonlinear interactions—makes deterministic earthquake prediction an elusive goal. Instead, statistical and thermodynamic approaches have emerged as powerful tools for understanding seismic phenomena.

Among these, Non-Extensive Statistical Mechanics (NESM), introduced by Tsallis (1988) \cite{Tsallis1988,Tsallis2011}, provides a robust framework for describing systems with long-range interactions, multifractality, and strong correlations—features intrinsic to seismogenesis. The fragment-asperity model, which conceptualizes fault zones as a collection of irregular, stress-bearing fragments (asperities) embedded in a viscoelastic matrix, aligns naturally with NESM, offering a way to describe the statistical distribution of earthquake magnitudes and energy release 

The analysis of seismic events using NESM, particularly Tsallis entropy, has provided significant insights into the complex dynamics of earthquakes. Several studies have demonstrated the presence of non-extensivity in earthquake magnitude and interevent time distributions, supporting the applicability of Tsallis statistics in seismology.

Silva \textit{et al}. \cite{Silva2006} and Vilar \textit{et al}. \cite{Vilar2007} were among the first to apply NESM to seismic catalogs, showing that earthquake magnitude distributions deviate from the traditional Gutenberg-Richter law and are better described by a $q$-exponential distribution. In subsequent years, several authors have reported on the NESM treatment in relation with different tectonic seismic events throughout the world \cite{Darooneh2009,Telesca2010-1,Telesca2010-2,Telesca2010-3,Vallianatos2010,Vallianatos2012,Telesca2011,Papadakis2015,Barahona2024,Sigalotti2023,Sigalotti2024}.

\begin{figurehere}
	\centering	\includegraphics[width=1\columnwidth,angle=0]{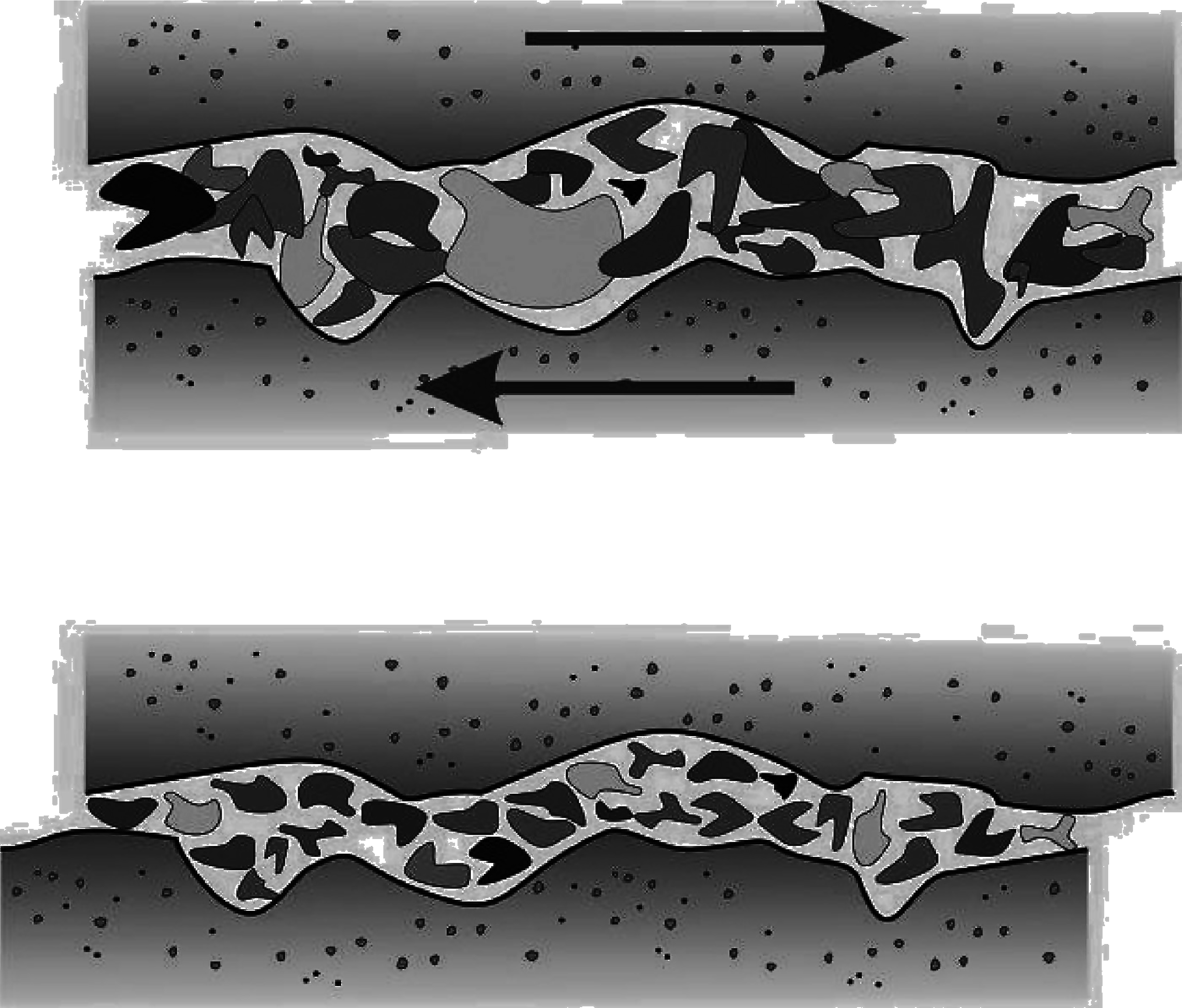}
	\caption{{\small (Top panel - Pre-seismic phase) Before an earthquake occurs, the system exhibits heterogeneous stress distribution and a broad spectrum of fragment sizes. (Bottom panel - Co-seismic phase) The earthquake rupture process leads to fragmentation of asperities and breakdown of barriers, resulting in size homogenization of the fractured materials.}}
	\label{Fig1}
\end{figurehere}

As pointed out by Barahona C\'ardenas and Araujo Soria, the complex system under study also encompasses all faults in the Earth's crust, which themselves exhibit intricate structural properties. By applying nonextensive statistical analysis to subduction zones, we can account for the diverse mechanisms driving seismic activity. Such systems are characterized by long-range spatiotemporal correlations, non-Markovian processes (indicating long-term memory effects), additive and multiplicative noise in mesoscopic Langevin-type equations, weak chaos with near-zero Lyapunov exponents, multifractal geometry, and long-range interactions in many-body systems. Additionally, they often involve nonlinear and/or inhomogeneous mesoscopic Fokker-Planck equations \cite{Barahona2024}. 

Within this context, the fragment-asperity model considers this complexity and describes the tectonic plates having numerous surface asperities with variable size and a whole distribution of fragments accommodating in between. It can be said that all asperities interact to each other via the stress propagation through the fragments. Therefore, this interaction has long range nature. In the relative motion of subduction certain asperities can break, giving origin to a seismic event. Sotolongo and Posadas put forward this idea in 2004 \cite{Sotolongo2004}, and extended it in 2023 \cite{Posadas2023}. Their proposal allowed to modify the Guttenberg-Richter (GR) law into one that includes the $q$-parameter. This makes possible to determine $q$ from the analysis of earthquake catalogs. Examples of the most recent studies along these lines are the references \cite{Barahona2024,Sigalotti2023,Sigalotti2024}. 

Essentially, the fragment-asperity model relies on the following physical picture: Before an earthquake occurs, the system exhibits heterogeneous stress distribution and a broad spectrum of fragment sizes. This configuration allows for numerous possible microscopic configurations ("microstates"), corresponding to a state of relatively high entropy. Then, the earthquake rupture process leads to fragmentation of asperities and breakdown of barriers, resulting in size homogenization of the fractured materials. This reduction in geometric variability decreases the number of allowable microstates, causing a sudden drop in entropy. The entropy reduction occurs rapidly during the dynamic rupture phase, followed by gradual recovery as tectonic stresses begin to rebuild in the post-seismic period \cite{Posadas2023}. A schematic representation of the fragment-asperity situation in a tectonic seismic phenomenon appears in Fig. 1.


In this communication, we provide an explanation for the above-mentioned interval of criticality for the $q$-parameter reported in many situations linked to tectonic earthquakes. This is done within the fragment-asperity model by making the novel assumption that the energy is, actually, proportional to the surface of interacting elements (asperities and fragments). Then, the calculation of total entropy, $S$, leads to a functional dependence of this quantity versus $q$ in which it is straightforward to identify the role of this parameter by maximizing $S(q)$. We are aimed at discussing the possible role of $q$-value as a main player in the prospective elaboration of a predictive system for tectonic earthquakes.




\section{Theoretical Framework}\label{theory}


The main assumption within the non-extensive treatment of the fragment-asperity model \cite{Sotolongo2004} is that the released seismic energy, $\varepsilon$ is directly related to the size of the fragments that occupy the space between fault blocks in tectonic plates: $\varepsilon\sim r$. The revision made by Silva \textit{et al} put forward a volume-related relationship $\varepsilon\sim r^3$. In the present study we propose that the seismic energy is, actually, proportional to the contact surface of the fragments: $\varepsilon\sim \sigma$. One may realize by observing the schematic representation in the upper panel Fig. 1, that strain in the system accumulates not only via the contact sites between the very fragments but also through those between fragments and plates. Each different configuration can be considered as a compatible "microstate" \cite{Posadas2023}. As it is known, entropy has to do with the multiplicity of these "micro" configurations. It tends to be maximal when there are many more ways of stress contact between fragments and plate asperities, which associates with a non-uniform distribution of fragment sizes. Clearly, such a contact directly relates with the surface of the fragment that undergoes the stress from its surroundings. Therefore, one may construct the probability distribution as a function of the area $\sigma$. In a continuous view, the reduced ($k=1$) Tsallis entropy 

\begin{equation}
	S=\frac{1-\int_{0}^{\infty}p^q(\sigma)d\sigma}{q-1},
\end{equation}

\noindent subject to the constraints of normalization,

\begin{equation}
\int_{0}^{\infty}p(\sigma)d\sigma=1,
\end{equation}

\noindent and $q$-mean of the distribution, 

\begin{equation}
\int_{0}^{\infty}\sigma p^q(\sigma)d\sigma=\langle\langle \sigma\rangle\rangle_q <\infty,
\end{equation}

\noindent is maximized to give the fragment size distribution

\begin{equation}
	p(\sigma)=\frac{(2-q)^{\frac{1}{2-q}}}{\left[1+(q-1)(2-q)^{\frac{q-1}{2-q}}\sigma\right]^{\frac{1}{q-1}}}.
\end{equation}

Now, we calculate the total entropy by integrating in (1) with respect to the surface $\sigma$, using (4). This gives

\begin{equation}
	S(q)=\frac{1-(2-q)^{\frac{1}{2-q}}}{q-1}.
\end{equation}

It is of particular interest to represent the plots of $S(q)$ and its derivative with respect to $q$. They are shown in Fig. 2.

\begin{widetext}
\begin{figurehere}
	\centering
	\includegraphics[width=0.45\columnwidth,angle=0]{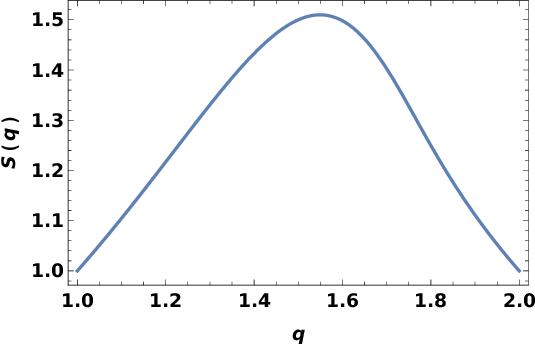}
	\includegraphics[width=0.45\columnwidth,angle=0]{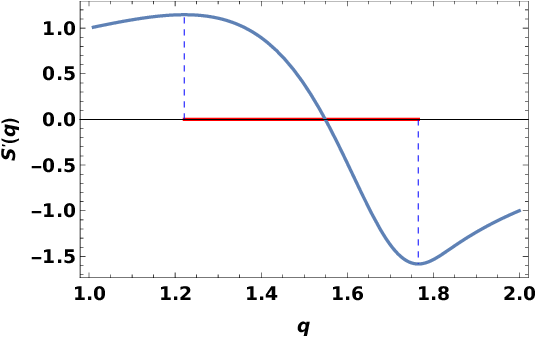}	
	\caption{(Left) Total non-extensive (reduced) entropy, $S(q)$, of the fragment-asperity model as a function of the $q$-parameter. (Right) The first derivative of $S(q)$, as a function of the $q$-parameter. The marked horizontal interval extends between the first ($q=1.22$) and second ($q=1.76$) inflection points of the entropy. It is readily apparent that the rate of change of $S(q)$ is more pronounced in the range of $1.4\leq q\leq 1.75$.  }
    \label{Fig2}
\end{figurehere}
\end{widetext}




\section{Results and discussion}\label{results}


We aim to analyze the outcome of this approach considering the different reported values of the non-extensivity parameter in studies applying NESM to the seismic catalogs referenced in the introduction. For instance, in the early works of Refs. \cite{Silva2006,Vilar2007}, $q$-parameter values lie typically in the range $1.6 \leq q \leq 1.7$, indicating strong non-extensive behavior.

Further studies by Darooneh and Mehri \cite{Darooneh2009} and Telesca \cite{Telesca2010-1,Telesca2010-2,Telesca2010-3} reinforced these findings. Telesca \cite{Telesca2010-1} examined Italian seismicity, finding $q \approx 1.6–1.8$, suggesting long-range interactions and memory effects in earthquake dynamics. In another study \cite{Telesca2010-2}, he analyzed seismic sequences in different tectonic settings, obtaining $q \approx 1.5–1.9$, with higher $q$-values correlating with more complex fault systems.

Vallianatos and Sammonds \cite{Vallianatos2010} and Vallianatos \textit{et al}. \cite{Vallianatos2012} expanded this framework to different regions, including the San Andreas Fault, reporting $q \approx 1.6–1.8$, consistent with criticality in earthquake preparation processes. Telesca further confirmed these results, showing that the $q$-parameter varies with tectonic activity, with higher values ($q > 1.7$) in regions of intense deformation \cite{Telesca2011}.

These studies collectively demonstrate that seismic phenomena exhibit non-extensive behavior, characterized by $q$-values consistently greater than 1 (the Boltzmann-Gibbs limit), reflecting the presence of long-range correlations, multifractality, and scale-invariant properties in earthquake dynamics. The $q$-parameter thus serves as a key indicator of the degree of non extensivity in seismic systems, providing deeper understanding of their underlying physics. Reports about the increase of $q$ previous the occurrence of earthquakes of high magnitude have appeared in recent times \cite{Papadakis2015,Barahona2024}.

Clearly, the range of values of $q$-parameter identified in the vast majority of cases coincides with the interval of most pronounced variation in Tsallis entropy within the fragment-asperity model, as can be seen in the right panel of Fig. 2. This interval includes the maximum value of this quantity as a function of q, which is $q_m = 1.55$.

If we consider, for example, the most recent reports (2024), we find that in the analysis of seismic activity along Mexico's western coastline (recorded in the 2000-2023 catalog of Mexican National Seismological Service), the general result for the non-extensivity parameter is $1.52 \lesssim q \lesssim 1.61$ \cite{Sigalotti2024}. This activity is primarily associated with tectonic plate interactions along the southern Pacific coast as well as active structures in the Gulf of California. Meanwhile, the study of data from Ecuador's subduction zone (1901-2021) shows that the occurrence of $M > 5$ earthquakes is associated with values of $1.68 \leq q\leq 1.75$, with noticeable decreases in this parameter following such events \cite{Barahona2024}. One interesting fact revealed in this latter study is the relation of $q$ versus time around the date of the $M=7.8$ Manab\'i earthquake of April 2016: A sudden increase just before the event, followed by a pronounced fall after it. This fact coincides with the observations of Papadakis \textit{et al }. for the Kobe earthquake of 1995 \cite{Papadakis2015} and Telesca for the L'Aquila one in 2009 \cite{Telesca2010-3}. 

Those many elements of agreement significantly correlate with what we see in Fig. 2 for $S(q)$ and $S^{\prime}(q)$. As a function of the complexity, represented by the $q$-parameter, the interval of highest entropy values and of its steepest variation, coincides with the range arising from the analysis of plate subduction seismicity around the world -largely in relation with mainshocks ($M>5$). This mathematical feature speaks about the criticality related with tectonic activity, described using NESM within the fragment-asperity model. Also, it points at confirming the long-range nature of the involved interactions as well as the need for a non-extensive description of these phenomena.

In our opinion, to identify the role of $q$ in the properties of total Tsallis entropy as a measure of the microstates distribution in the statistical approach to tectonic seismicity becomes a conceptual breakthrough that strengthens the fragment-asperity model and definitively removes this number from the category of fitting parameter, as it has been considered in previous works. At the same time, our analysis theoretically supports considering $q$ as an indicator of criticality.

Now, we can ask ourselves: How can this result be useful to earthquake prediction? 

First and foremost, by assuming the direct dependence of seismic energy on the contact surface between fragments and asperities we have obtained a suitable tool to identify a range of critical values when dealing with the seismic dynamics: the total Tsallis entropy $S(q)$ [eqn (5)].

The $q$-parameter now carries more concrete significance: it serves not merely as a measure of proximity to extensivity, but also as a potential indicator of seismic risk in a given region. Current methodologies enable analysis of major earthquakes by comparing their $q$-values with preceding $q$-sequences, offering insights into predictive precision. However, standalone $q$-values or their temporal sequences remain insufficient for exact predictions —we cannot yet determine the precise timing or location of seismic events. Nevertheless, identifying critical $q$-value ranges warrants focused research attention, as $q$-sequence analysis from seismic catalogs may yield valuable information.
	
	 
At present, to make exact earthquake forecasting lies beyond our capabilities. Together with NESM approach it would be necessary, among other elements: (i) to deploy denser seismic networks aiming at better resolving the subsurface dynamics; (ii) to implement higher-resolution temporal sampling;  (iii) developing novel monitoring approaches. Moreover, to improve prediction, $q$ must be used together with other indicators such  as measuring the acceleration of seismic activity and determination of changes in the GR $b$-relation. In all this, the use of machine learning techniques (e.g. see Refs. \cite{Liang2023,Convertito2024}) may signify enabling high-resolution spatiotemporal analysis (e.g., identifying critical subregions or optimizing temporal windows).

	

%


\section{Conclusions}\label{conclusions}

The possibility of evaluating the total Tsallis entropy as a function of the non-extensivity parameter for the fragment-asperity model allows to identify the interval of $q$-values where this quantity exhibits its more pronounced variation, including its maximum value. It is shown that such interval includes the ranges determined for $q$ in all investigated NESM approaches of tectonic seismic activity around the world. This fact directly speaks about the complexity of this natural phenomenon, highlights the particular role of $q$ beyond its use as a fitting parameter, and could become a relevant tool in the construction of a predictive procedure for earthquakes. In conjunction with other elements discussed in this article, the surveillance of $q$-values evolution would allow to identify the proximity of criticality in tectonic plate interaction within a given subregion and to alert about the possible occurrence of a seismic event.

As a case study, the controversial fracking process can be investigated through $q$-value dynamics in active extraction zones. Such analysis could disentangle the relative contributions of different mechanisms (e.g. plate fracturing, fluid injection), quantify their environmental impacts, particularly regarding induced seismicity and establish causal hierarchies among triggering factors. A manuscript addressing this specific application is currently in preparation.

\bibliography{bibliography-Sotolongo}

\begin{thebibliography}{19}
\expandafter\ifx\csname natexlab\endcsname\relax\def\natexlab#1{#1}\fi
\expandafter\ifx\csname bibnamefont\endcsname\relax
  \def\bibnamefont#1{#1}\fi
\expandafter\ifx\csname bibfnamefont\endcsname\relax
  \def\bibfnamefont#1{#1}\fi
\expandafter\ifx\csname citenamefont\endcsname\relax
  \def\citenamefont#1{#1}\fi
\expandafter\ifx\csname url\endcsname\relax
  \def\url#1{\texttt{#1}}\fi
\expandafter\ifx\csname urlprefix\endcsname\relax\def\urlprefix{URL }\fi
\providecommand{\bibinfo}[2]{#2}
\providecommand{\eprint}[2][]{\url{#2}}

\bibitem[{\citenamefont{Tsallis}(1988)}]{Tsallis1988}
\bibinfo{author}{\bibfnamefont{C.}~\bibnamefont{Tsallis}},
  \bibinfo{journal}{Journal of Statistical Physics}
  \textbf{\bibinfo{volume}{52}}, \bibinfo{pages}{479} (\bibinfo{year}{1988}).

\bibitem[{\citenamefont{Tsallis}(2011)}]{Tsallis2011}
\bibinfo{author}{\bibfnamefont{C.}~\bibnamefont{Tsallis}},
  \bibinfo{journal}{Entropy} \textbf{\bibinfo{volume}{13}},
  \bibinfo{pages}{1765} (\bibinfo{year}{2011}).

\bibitem[{\citenamefont{Silva et~al.}(2006)\citenamefont{Silva,
  Fran\ifmmode~\mbox{\c{c}}\else \c{c}\fi{}a, Vilar, and Alcaniz}}]{Silva2006}
\bibinfo{author}{\bibfnamefont{R.}~\bibnamefont{Silva}},
  \bibinfo{author}{\bibfnamefont{G.~S.}
  \bibnamefont{Fran\ifmmode~\mbox{\c{c}}\else \c{c}\fi{}a}},
  \bibinfo{author}{\bibfnamefont{C.~S.} \bibnamefont{Vilar}}, \bibnamefont{and}
  \bibinfo{author}{\bibfnamefont{J.~S.} \bibnamefont{Alcaniz}},
  \bibinfo{journal}{Phys. Rev. E} \textbf{\bibinfo{volume}{73}},
  \bibinfo{pages}{026102} (\bibinfo{year}{2006}).

\bibitem[{\citenamefont{Vilar et~al.}(2007)\citenamefont{Vilar, França, Silva,
  and Alcaniz}}]{Vilar2007}
\bibinfo{author}{\bibfnamefont{C.}~\bibnamefont{Vilar}},
  \bibinfo{author}{\bibfnamefont{G.}~\bibnamefont{França}},
  \bibinfo{author}{\bibfnamefont{R.}~\bibnamefont{Silva}}, \bibnamefont{and}
  \bibinfo{author}{\bibfnamefont{J.}~\bibnamefont{Alcaniz}},
  \bibinfo{journal}{Physica A: Statistical Mechanics and its Applications}
  \textbf{\bibinfo{volume}{377}}, \bibinfo{pages}{285} (\bibinfo{year}{2007}).

\bibitem[{\citenamefont{Darooneh and Mehri}(2010)}]{Darooneh2009}
\bibinfo{author}{\bibfnamefont{A.~H.} \bibnamefont{Darooneh}} \bibnamefont{and}
  \bibinfo{author}{\bibfnamefont{A.}~\bibnamefont{Mehri}},
  \bibinfo{journal}{Physica A: Statistical Mechanics and its Applications}
  \textbf{\bibinfo{volume}{389}}, \bibinfo{pages}{509} (\bibinfo{year}{2010}).

\bibitem[{\citenamefont{Telesca}(2010{\natexlab{a}})}]{Telesca2010-1}
\bibinfo{author}{\bibfnamefont{L.}~\bibnamefont{Telesca}},
  \bibinfo{journal}{Tectonophysics} \textbf{\bibinfo{volume}{494}},
  \bibinfo{pages}{155} (\bibinfo{year}{2010}{\natexlab{a}}).

\bibitem[{\citenamefont{Telesca}(2010{\natexlab{b}})}]{Telesca2010-2}
\bibinfo{author}{\bibfnamefont{L.}~\bibnamefont{Telesca}},
  \bibinfo{journal}{Physica A: Statistical Mechanics and its Applications}
  \textbf{\bibinfo{volume}{389}}, \bibinfo{pages}{1911}
  (\bibinfo{year}{2010}{\natexlab{b}}).

\bibitem[{\citenamefont{Telesca}(2010{\natexlab{c}})}]{Telesca2010-3}
\bibinfo{author}{\bibfnamefont{L.}~\bibnamefont{Telesca}},
  \bibinfo{journal}{Terra Nova} \textbf{\bibinfo{volume}{22}},
  \bibinfo{pages}{87} (\bibinfo{year}{2010}{\natexlab{c}}).

\bibitem[{\citenamefont{Vallianatos and Sammonds}(2010)}]{Vallianatos2010}
\bibinfo{author}{\bibfnamefont{F.}~\bibnamefont{Vallianatos}} \bibnamefont{and}
  \bibinfo{author}{\bibfnamefont{P.}~\bibnamefont{Sammonds}},
  \bibinfo{journal}{Physica A: Statistical Mechanics and its Applications}
  \textbf{\bibinfo{volume}{389}}, \bibinfo{pages}{4989} (\bibinfo{year}{2010}).

\bibitem[{\citenamefont{Vallianatos et~al.}(2012)\citenamefont{Vallianatos,
  Benson, Meredith, and Sammonds}}]{Vallianatos2012}
\bibinfo{author}{\bibfnamefont{F.}~\bibnamefont{Vallianatos}},
  \bibinfo{author}{\bibfnamefont{P.}~\bibnamefont{Benson}},
  \bibinfo{author}{\bibfnamefont{P.}~\bibnamefont{Meredith}}, \bibnamefont{and}
  \bibinfo{author}{\bibfnamefont{P.}~\bibnamefont{Sammonds}},
  \bibinfo{journal}{Europhysics Letters} \textbf{\bibinfo{volume}{97}},
  \bibinfo{pages}{58002} (\bibinfo{year}{2012}).

\bibitem[{\citenamefont{Telesca}(2011)}]{Telesca2011}
\bibinfo{author}{\bibfnamefont{L.}~\bibnamefont{Telesca}},
  \bibinfo{journal}{Entropy} \textbf{\bibinfo{volume}{13}},
  \bibinfo{pages}{1267} (\bibinfo{year}{2011}).

\bibitem[{\citenamefont{Papadakis et~al.}(2015)\citenamefont{Papadakis,
  Vallianatos, and Sammonds}}]{Papadakis2015}
\bibinfo{author}{\bibfnamefont{G.}~\bibnamefont{Papadakis}},
  \bibinfo{author}{\bibfnamefont{F.}~\bibnamefont{Vallianatos}},
  \bibnamefont{and} \bibinfo{author}{\bibfnamefont{P.}~\bibnamefont{Sammonds}},
  \bibinfo{journal}{Pure and Applied Geophysics}
  \textbf{\bibinfo{volume}{172}}, \bibinfo{pages}{1923} (\bibinfo{year}{2015}).

\bibitem[{\citenamefont{Barahona~Cárdenas and
  Araujo~Soria}(2024)}]{Barahona2024}
\bibinfo{author}{\bibfnamefont{D.~A.} \bibnamefont{Barahona~Cárdenas}}
  \bibnamefont{and} \bibinfo{author}{\bibfnamefont{J.~S.}
  \bibnamefont{Araujo~Soria}}, \bibinfo{journal}{Geofísica Internacional}
  \textbf{\bibinfo{volume}{63}}, \bibinfo{pages}{1165–1174}
  (\bibinfo{year}{2024}).

\bibitem[{\citenamefont{Sigalotti et~al.}(2023)\citenamefont{Sigalotti,
  Ramírez-Rojas, and Vargas}}]{Sigalotti2023}
\bibinfo{author}{\bibfnamefont{L.~D.~G.} \bibnamefont{Sigalotti}},
  \bibinfo{author}{\bibfnamefont{A.}~\bibnamefont{Ramírez-Rojas}},
  \bibnamefont{and} \bibinfo{author}{\bibfnamefont{C.~A.}
  \bibnamefont{Vargas}}, \bibinfo{journal}{Entropy}
  \textbf{\bibinfo{volume}{25}} (\bibinfo{year}{2023}).

\bibitem[{\citenamefont{Flores-Márquez
  et~al.}(2024)\citenamefont{Flores-Márquez, Ramírez-Rojas, and
  Sigalotti}}]{Sigalotti2024}
\bibinfo{author}{\bibfnamefont{E.~L.} \bibnamefont{Flores-Márquez}},
  \bibinfo{author}{\bibfnamefont{A.}~\bibnamefont{Ramírez-Rojas}},
  \bibnamefont{and} \bibinfo{author}{\bibfnamefont{L.~D.~G.}
  \bibnamefont{Sigalotti}}, \bibinfo{journal}{Fractal and Fractional}
  \textbf{\bibinfo{volume}{8}} (\bibinfo{year}{2024}).

\bibitem[{\citenamefont{Sotolongo-Costa and Posadas}(2004)}]{Sotolongo2004}
\bibinfo{author}{\bibfnamefont{O.}~\bibnamefont{Sotolongo-Costa}}
  \bibnamefont{and} \bibinfo{author}{\bibfnamefont{A.}~\bibnamefont{Posadas}},
  \bibinfo{journal}{Phys. Rev. Lett.} \textbf{\bibinfo{volume}{92}},
  \bibinfo{pages}{048501} (\bibinfo{year}{2004}).

\bibitem[{\citenamefont{Posadas and Sotolongo-Costa}(2023)}]{Posadas2023}
\bibinfo{author}{\bibfnamefont{A.}~\bibnamefont{Posadas}} \bibnamefont{and}
  \bibinfo{author}{\bibfnamefont{O.}~\bibnamefont{Sotolongo-Costa}},
  \bibinfo{journal}{Communications in Nonlinear Science and Numerical
  Simulation} \textbf{\bibinfo{volume}{117}}, \bibinfo{pages}{106906}
  (\bibinfo{year}{2023}).

\bibitem[{\citenamefont{Liang et~al.}(2023)\citenamefont{Liang, Xu, Chen, and
  Jiang}}]{Liang2023}
\bibinfo{author}{\bibfnamefont{X.}~\bibnamefont{Liang}},
  \bibinfo{author}{\bibfnamefont{T.}~\bibnamefont{Xu}},
  \bibinfo{author}{\bibfnamefont{J.}~\bibnamefont{Chen}}, \bibnamefont{and}
  \bibinfo{author}{\bibfnamefont{Z.}~\bibnamefont{Jiang}},
  \bibinfo{journal}{Renewable Energy} \textbf{\bibinfo{volume}{216}},
  \bibinfo{pages}{119046} (\bibinfo{year}{2023}).

\bibitem[{\citenamefont{Convertito et~al.}(2024)\citenamefont{Convertito,
  Giampaolo, Amoroso, and Piccialli}}]{Convertito2024}
\bibinfo{author}{\bibfnamefont{V.}~\bibnamefont{Convertito}},
  \bibinfo{author}{\bibfnamefont{F.}~\bibnamefont{Giampaolo}},
  \bibinfo{author}{\bibfnamefont{O.}~\bibnamefont{Amoroso}}, \bibnamefont{and}
  \bibinfo{author}{\bibfnamefont{F.}~\bibnamefont{Piccialli}},
  \bibinfo{journal}{Scientific Reports} \textbf{\bibinfo{volume}{14}},
  \bibinfo{pages}{2964} (\bibinfo{year}{2024}).

\end{thebibliography}
\end{document}